\documentclass{article}       
\usepackage{amssymb}
\usepackage{graphicx,graphics}
\usepackage{amsmath}
\usepackage[shortlabels]{enumitem}
\usepackage[final]{changes}
 \usepackage{verbatim} 

\usepackage[export]{adjustbox}
\usepackage{tikz}
\usepackage{url}
\usepackage{multirow}
\usepackage{setspace} 
\usepackage{booktabs}
\usepackage{subcaption}
\usepackage{rotating}

\usepackage[affil-it]{authblk}

\begin{document}
	
	\title{Environmentally extended input-output analysis in complex networks: a multilayer approach}
	\author[1]{Alessandra Cornaro \thanks{alessandra.cornaro@unimib.it}}
	\author[1]{Giorgio Rizzini \thanks{giorgio.rizzini@unimib.it}}

	\affil[1]{Universit\`{a} degli Studi di Milano-Bicocca - Dipartimento di Statistica e Metodi Quantitativi, via Bicocca degli Arcimboldi, 8 - 20216 Milano MI - Italy}

\date{}

\maketitle

	\begin{abstract}
In this paper we propose a methodology suitable for a comprehensive analysis of the global embodied energy flow trough a complex network approach. To this end, we extend the existing literature, providing a multilayer framework based on the environmentally extended input-output analysis. The multilayer structure, with respect to the traditional approach, allows us to unveil the different role of sectors and economies in the system. In order to identify key sectors and economies, we make use of hub and authority scores, by adapting to our framework an extension of the Kleinberg algorithm, called Multi-Dimensional HITS (MD-HITS). A numerical analysis based on multi-region input-output tables shows how the proposed approach provides meaningful insights. 

	\end{abstract}


	\section{Introduction}
	\label{intro}
	
The human impact on the natural environment continues to grow together with an important resources’ depletion, in particular energy assets.
The availability of abundant and affordable energy is indeed a key feature of modern and prosperous societies but the uneven distribution of energy resources gave rise to an increasing international energy commodities trade with the aim to satisfy the energy demand of those countries not able to rely on energy domestic production. Furthermore, due to the globalization, the transfer of direct and indirect energy resources through international trade, has intensified. Additionally, current and future scenarios are characterized by a transition towards increasing investments in renewable energies. The world has indeed committed to decarbonisation, but the pathways how to get there are still uncertain. \\
Usually, a common practice to assess energy consumption is related to direct energy. However, the production of goods and services consumes not only direct energy, like coal, gas, oil etc., but also indirect energy. Indirect energy (or embodied energy) is the energy consumption embodied in goods and services production to make the final product (see, e.g., \cite{Brown} and \cite{qier}).
Therefore, focusing only on direct energy trade does not allow to catch completely the complexity and the full range of energy transfer among economies. To tackle this issue and to provide a holistic and systematic framework of energy use, it is indeed important to address our attention to the global energy (indirect and direct) involved in the production of goods and services.\\
In the literature, input-output analysis has been intensively exploited with the aim to quantify energy embodied in trade at both national and international level (see \cite{Miller}, \cite{Sun} and \cite{Wiedmann2}).
The classical input-output model was originally developed by Leontief (see \cite{Leontief1} and \cite{Leontief2}) in order to analyse the interdependency of all sectors of the economy and how the production of an output in one industry or sector can affect the national economy.
Later, the basic methodology for input–output analysis evolved, in both theory and applications, and it has been extended by a variety of contributions aimed at evaluating the environmental and social impact of the economic activities. 
In this fashion, the environmentally extended input-output analysis (EEIOA) (see, among others, \cite{Kites}, \cite{Miller} and \cite{Wiedmann})  has been developed and it represents a comprehensive method for evaluating the linkages between economic consumption activities and environmental impacts, including the harvest and degradation of natural resources.
In such a framework, the interconnected and intricate embodied energy flows in international trade have been explored from a network perspective in a variety of studies. Many network tools have been implemented in order to explore the structure of the embodied energy flows in trade at different scales as global, regional and national levels and to analyse various direct energy commodity trade network (see, among others, \cite{Chen2}, \cite{Du}, \cite{Hao} and \cite{Zhong}). It is noteworthy to say that these contributions focus only on monoplex networks, where the countries/economies are treated as nodes and embodied energy flows as edges. However, the monoplex structure does not provide a comprehensive analysis of the global embodied energy flow because it accounts only for the role of an economy in the complex flow system while it is also relevant to highlight which are the key sectors in the system.
In order to tackle this issue, in this work we extend these previous studies, in particular the work of \cite{Chen2}, by building a multilayer network, the Embodied Energy Flow Multilayer Network (EEFMN), based on EEIOA data.  This new approach represents an effective tool to model the global embodied energy flow and to present a full picture of the whole complexity of the system. In particular, while traditional approaches tend to aggregate flows considering only countries, a multilayer structure allows us to unveil the different role of sectors and economies. To this end, by extending the methodology provided in \cite{Chen2}, we compute the embodied energy flows for each couple of economies and sectors. Energy flows are then used to construct a directed and weighted multilayer network where each layer represents the economy/country\footnote{We shall use industries and sectors as synonyms. Moreover, economies and countries will be used as synonyms.}. In each layer, sectors/industries are nodes and arcs take into account directed intra-layer embodied energy flows between different sectors of the same economy. Inter-layer arcs consider instead directed energy flows between the same sector in different countries or different sectors in different countries. In this way, with respect to the existing literature, we are able to consider both the role of economies and sectors in the network. \\
Secondly, we use centrality measures in order to identify economies and sectors that play a key role acting as collectors and distributors in global energy flow system. A suitable measure in this context is represented by hub and authority scores provided by Kleinberg in \cite{Kleinberg1999}. Hence, we adapt to our framework an extension of the Kleinberg algorithm, called Multi-Dimensional HITS (MD-HITS), developed by \cite{Arrigo}. In this generalization for multilayer networks, not only nodes are ranked but also layers receive a score, ranking their capacity of linking important layers or being linked by hub layers. Hence, including layers to be used to compute centrality, the model defines five centrality vectors: two for nodes (hub and authority scores), two for layers (broadcasting and receiving scores) and one for temporal evolution\footnote{It is worth pointing out that, differently from \cite{Arrigo}, the fifth score related to the temporal aspect is here not considered since data refer to a specific calendar year and a temporal analysis is out of the scope of this paper.} Also the proposed measures are new in this context, since embodied energy flows have been studied only using classical centrality measures based on monoplex networks. \\
The proposed approach has been applied to a multilayer network of 189 countries and 26 sectors collecting data from a global supply chain database\footnote{See https://worldmrio.com/} in 2016. It consists in multi-region input-output tables (MRIO), that provide a time series of high-resolution input-output tables with matching environmental and social satellite accounts. The analysis of the topological structure of EEFMN is suitable to detect features, evolutionary characteristics and the relationships among sectors and economies. The additional insights given by the multilayer structure help us to identify not only the different roles of economies in the complex flow system but even which are peculiar sectors. Numerical results confirm that relevant countries in the world trade have also a central role in terms of energy consumption. This result is in line with main patterns identified in \cite{Chen2}. Additionally, the procedure allows to highlight relevant sectors that generate energy flows and sectors that act as collectors consuming large volumes of energy for powering buildings and tools. \\
Previous results can be important also in terms of policy implications. The methodology allows to deeply understand the structure of countries economy highlighting their main strategic sectors in terms of both economic trade values and energy flows, that can be seen as a proxy of the environmental damages. Moreover, the approach based on a directed multilayer network can shed light on the power of each country in a specific sector and, therefore, assess how each country can affect the market structure. The latter aspect is particularly important to identify the results, in terms of economic trade values, power and environmental features, of each country’s economic policies, as, for instance, the decision to block the export of a certain product. \\
The connection through a multilayer network between economic trade and energy flows is a useful tool in the hands of any policymaker. With this approach, a policymaker can evaluate the outcomes of any economic and/or environmental policy. In particular, the multilayer approach allows to see how the world economic structure reacts to an external shock. For instance, one country's environmental policies, such as raising taxes on the most polluting sectors, may have an impact on the other countries. In this context, the use of a multilayer network helps us to identify at a global level which are the strategic sectors for the achievement of environmental goals, also taking into account the economic dimension of these sectors. In doing so, it is possible to establish environmental strategies that considering the trade-off between economic activity and the environment, may ensure a sustainable economy growth.\\
The paper is organized as follows. In Section \ref{sec:Pre} we introduce some preliminaries and notations useful to understand the proposed approach. In Section \ref{sec:meth} we illustrate the methodology we developed in order to build the global embodied energy flow multilayer network, based on the environmentally extended input-output analysis. Furthermore, we present centrality measures aimed at identifying economies and sectors that play a key role acting as collectors and distributors in the global energy flow system. Numerical analysis is reported in Section \ref{sec:na}. Conclusions follow.
	
	\section{Preliminaries}
	\label{sec:Pre}
	
	In this section we recall some definitions and concepts about directed and weighted graph (see, among others, \cite{Harary}) useful to introduce the notations we will use throughout the paper.
	
	A directed and weighted network is a triple $G=(V,E,w)$, where $V$ is a set of $N$ vertices (or nodes), $E$ is the set of the ordered pairs of elements of $V$ (arcs or directed edges) and a real positive weight $w$ is assigned to each arc. 
	Two nodes are adjacent if there is at least an arc $(i,j)$ from node $i$ to $j$. 

	A weighted directed network is completely described by the real $N$-square matrix $\textbf{W}^{[\alpha]}$, the weighted adjacency matrix, whose entries are $w^{[\alpha]}_{ij}\neq0$ if there is an arc $(i,j)\in E$, and $w^{[\alpha]}_{ij}= 0$ otherwise.
	We make use of the notation $\textbf{W}^{[\alpha]}$ for the weighted adjacency matrix to be consistent with the general notation of the weighted supradjacency matrix defined later for multilayer networks.

	We now introduce the concept of strength centrality by using the same notation. We define the in-strength $s^{[\alpha]}_{i,in}$ and the out-strength $s^{[\alpha]}_{i,out}$ of a node $i$ as follows:

\begin{equation}
	s^{[\alpha]}_{i,in}=\left({\textbf{W}^{[\alpha]}}^T \textbf{1} \right)_{i} 
\end{equation}

and
\begin{equation}
	s^{[\alpha]}_{i,out}=\left(\textbf{W}^{[\alpha]} \textbf{1}\right)_i,
\end{equation}
where $\left(\textbf{W}^{[\alpha]}\right)_i$ and $\left({\textbf{W}^{[\alpha]}}^T\right)_{i}$ are respectively the $i$-th row of the weighted adjacency matrix and its transpose, and $\textbf{1}$ is the unit column vector of $N$ elements.

\noindent The total strength of $i$ is then:
\begin{equation}\label{tot_str}
	s^{[\alpha]}_{i,tot}=s^{[\alpha]}_{i,in}+s^{[\alpha]}_{i,out} .
\end{equation}

	We then define the eigenvector centrality, another important centrality measure. For an undirected weighted graph, the eigenvector centrality of a node $i$ is the $i$-th component of the eigenvector associated with the maximum eigenvalue of the weighted adjacency matrix (see \cite{Bonacichb} and \cite{Bonacicha}). The extension of this measure to directed networks leads to the definition of two scores, hub and authority, associated with a node. The concept has been introduced by \cite{Kleinberg1999} to rank the importance of a web page. Let $x^{[\alpha]}_i$ and $y^{[\alpha]}_i$ be the hub and authority scores for node $i$, respectively.
	They are the $i$-th component of $\textbf{x}$ and $\textbf{y}$ that collect the hubs and authority scores, respectively, of all nodes. These vectors are obtained by the following relations:
	\begin{equation}\label{auth}
		\textbf{x}^{[\alpha]}={\textbf{W}^{[\alpha]}}^T\textbf{y}^{[\alpha]}
	\end{equation}
	and 
	\begin{equation}\label{hubs}
		\textbf{y}^{[\alpha]}=\textbf{W}^{[\alpha]}\textbf{x}^{[\alpha]}.
	\end{equation}
 Hubs and authorities are characterized by a mutually reinforcing relationship, i.e., a good hub is a node that points to many good authorities; a good authority is a node to which many good hubs point. For computing the scores, an iterative algorithm (HITS - Hyperlink Induced Topic Search) is proposed (for details see \cite{Kleinberg1999}). 	 \\
	
We focus now on multilayer networks and, in particular, we consider node-aligned networks with non-diagonal couplings  (see \cite{Kivela}). A multilayer
	network consists of a family of networks $G_\alpha=(V_\alpha,E_{\alpha,\beta})$,  $\alpha,\beta=1,...,L$, where each network $G_\alpha=(V_\alpha,E_{\alpha,\beta})$ is located in a layer $\alpha$ and a node $i \in V_\alpha$ is adjacent to $j \in V_\beta$, $\forall \alpha,\beta=1,...,L$ if there is an arc connecting them. In node-aligned networks, nodes are the same over all layers, namely, $V_\alpha=V_\beta=V,\ \forall \alpha,\beta=1,...,L$, and
	$E_{\alpha,\beta}$ collects all the arcs connecting nodes on layer $\alpha$ to nodes within the same layer (intra-layer connections) and to nodes on different layers (inter-layer connections). We assume that the network is non-diagonal coupled, that is inter-layer arcs are allowed also between different nodes in different layers.
	
	\noindent A weight $w^{\left[\alpha\beta\right]}_{ij} > 0$ is associated with an arc $(i,j)$ in $E_{\alpha \beta}$.
	Note that when $\alpha=\beta$ we intend that there is a weighted arc $w^{\left[\alpha\right]}_{ij} > 0$ between nodes $i$ and $j$ in the network $G_\alpha$.

	The weighted adjacency relations between pair of nodes can be described by the weighted supradjacency matrix (or simply supradjacency). It is defined as a matrix, with $L-$square blocks, each one of order $N$:
	
	\begin{equation}
		\label{supra}
		\mathbf{W}=
		\begin{bmatrix}
			\mathbf{W}^{\left[1\right]} & \mathbf{W}^{\left[12\right]} & \cdots & \mathbf{W}^{\left[1L\right]} \\
			\mathbf{W}^{\left[21\right]} & \mathbf{W}^{\left[2\right]} & \cdots & \mathbf{W}^{\left[2L\right]} \\
			\vdots & \vdots & \ddots & \vdots \\
			\mathbf{W}^{\left[L1\right]} & \mathbf{W}^{\left[L2\right]} & \cdots & \mathbf{W}^{\left[L\right]}
		\end{bmatrix},
	\end{equation}
	
	\noindent where the diagonal blocks represent the weighted adjacency matrix of each layer $\textbf{W}^{\left[\alpha \right]}$, $\alpha=1,...,L$, whereas the non-diagonal blocks $\textbf{W}^{\left[\alpha \beta\right]}$ represent the weighted adjacency relations between nodes on layers $\alpha$ and nodes on layer $\beta$. 
	
	\noindent The generic element of the matrix $	\mathbf{W}$ is denoted as $w_{hk}$ with $h,k=1,...,NL$, where 
	\begin{equation}
		h=N(\alpha-1)+i, k=N(\beta-1)+j.
		\label{eq:hk}
	\end{equation} 
	Notice that the indices $h,k$ identify the position in the supradjacency matrix $\mathbf{W}$ of the weight of the arc $(i,j) \in E_{\alpha,\beta}$ (i.e. $w^{\left[\alpha\beta\right]}_{ij}=w_{hk}$). Thus, from now on, we always assume relation (\ref{eq:hk}) between $h,k$ and $\alpha,\beta,i,j$. 
	
	The in- and out-strength of a node $i$ in a layer $\alpha$ can be defined as follows:
	\begin{equation}\label{in_str}
		s^{[\alpha]}_{i,in}=(\mathbf{W}^T\mathbf{1})_h,
	\end{equation}	
	\begin{equation}\label{in_out}
		s^{[\alpha]}_{i,out}=(\textbf{W}\textbf{1})_h.
	\end{equation}
	
	\noindent Therefore, the total strength  $s^{[\alpha]}_i$ of a node $i$ in the layer $\alpha$ is 
	\begin{equation}\label{str}
		s^{[\alpha]}_i=s^{[\alpha]}_{i,in}+s^{[\alpha]}_{i,out}=[(\textbf{W}^T+\textbf{W})\textbf{1}]_h.
	\end{equation}
	
\noindent 	It is then possible to assess the in-strength of a node $i$ with respect to all layers:
	\begin{equation}\label{totin_str}
		s_{i,in}=\sum_{\alpha=1}^{L}s^{[\alpha]}_{i,in}
	\end{equation}
  and the in-strength of all nodes in a layer $\alpha$:
	\begin{equation}\label{totin_strL}
	s^{[\alpha]}_{in}=\sum_{i=1}^{N}s^{[\alpha]}_{i,in}.
\end{equation}

	Out-strength and total strength with respect to either all layers or all nodes can be defined similarly.

	\section{Methodology and data}
	\label{sec:meth}
	
	\subsection{Global embodied energy flow in a multilayer network framework}
	
	We now illustrate the methodology we developed in order to build the global embodied energy flow multilayer network. Our aim is to consider a directed and weighted node-aligned multilayer network, where nodes are represented by sectors and layers by economies. Hence, sectors and economies are denoted by $i=1,\dots,N$ and $\alpha=1,\dots,L$, respectively. In this way, intra-layer connections consider flows between different sectors within the same economy, while inter-layer takes into account flows between the same or different sectors in different economies.

	The multilayer network is fully described by a weighted supradjacency matrix $\mathbf{W}$ of order $NL$, defined as in formula (\ref{supra}), that can be partitioned into blocks. The blocks that belong to the main diagonal represent the embodied energy flows between sectors related to the same economy while the non-diagonal blocks are referred to embodied energy flows between sectors of different economies. The entries $w_{hk}$, with $h$ and $k$ specified in formula \eqref{eq:hk}, of matrix $\mathbf{W}$ are given by the embodied energy flows estimated via the following methodology based on the environmentally extended input-output analysis. Input–output analysis provides a useful framework in order to trace energy use and other related characteristics such as environmental pollution, associated with interindustry activity. 
    In this regard, the environmentally extended input-output analysis was developed to tackle the issue between economic consumption activities and environmental impacts.
    In a similar fashion, energy flows embodied in trade take into account only goods and service trade for final use.
    Then, through the Leontief inverse matrix, the technical energy consumption of all considered sectors is assigned to the final use.
    In order to compute the embodied energy flow, let us consider in our model a couple of sectors $i,j$, with $i,j=1, \dots, N$ and a couple of economies  $\alpha, \beta$, with $\alpha, \beta=1, \dots, L$, respectively.\\
    We start by introducing the direct requirement coefficient matrix (or input coefficient matrix) $\mathbf{A}$, of order $NL$, whose elements are defined as follows:
    
    \begin{equation}
    	a_{hk}=\dfrac{u_{ij}^{\alpha \beta}}{o_{j}^{\beta}},
    \end{equation}
    where $u_{ij}^{\alpha \beta}$ is the intermediate use of sector $j$ in economy $\beta$ provided by sector $i$ in economy $\alpha$ and $o_{j}^{\beta}$ is the total output of sector $j$ in economy $\beta$.
    We remind here that relations between $h,k$ and $\alpha, \beta, i,j$ are given by relation (\ref{eq:hk}).
    
    Notice that the elements of the matrix $\mathbf{A}$ measures the ratio of input values of goods and services used in each sector to the sector’s corresponding total output.\\
    
  The Leontief inverse matrix is defined as:
    \begin{equation}
    	 \mathbf{L}=(\mathbf{I}-\mathbf{A})^{-1},
    \end{equation}
where  $\mathbf{I}$ is the identity matrix of order $NL$ and $\mathbf{A}$ is defined as above.

The Leontief inverse matrix displays the total amount of goods and services required for the production of one unit of output and it can be seen as the decomposition of the total impact on output into its constituent layers according to the number of production stages involved.
Hence, the element $l_{ij}$ measures the successive effects on the economy as a result of the initial increase in production of an economic sector. In other words, this represents the spillover effect which arises between the various sectors of an economy.\\

Now, we can define the embodied energy flow as:
\begin{equation}
	\label{qij_network}
	q_{ij}^{\alpha \beta}= \mathbf{c}_{i}^{T}\mathbf{l}_{ij}^{\alpha}  f_{j}^{\alpha \beta}=\left( \sum_{\epsilon=1}^{L}  c_{i}^{\epsilon} l_{ij}^{\epsilon\alpha} \right)  f_{j}^{\alpha \beta} ,
\end{equation}

where:
\begin{itemize}
\item a vector  $\mathbf{c}_{i}=\left[c_{i}^{\epsilon} \right]$, with $\epsilon=1, \dots, L$, that stands for the technical energy consumption by sector $i$ in each investigated economy; 

\item a vector  $\mathbf{l}_{ij}^{\alpha}=\left[l_{ij}^{\epsilon\alpha} \right]$,  with $\epsilon=1, \dots, L$, that is the entry of the Leontief inverse matrix  $\mathbf{L}$ for sectors $i$ and $j$ and economies $\epsilon$ and $\alpha$;

\item a scalar $f_{j}^{\alpha \beta}$ that represents the goods/services traded as a final demand by sector $j$ from economy $\alpha$ to economy $\beta$.

\end{itemize}

Value of $q_{ij}^{\alpha \beta}$ represents the total embodied energy that is needed from $i$ in order to produce the good/services that are traded in sector $j$ as a final demand from economy $\alpha$ to economy $\beta$.

Therefore, we are able to identify the elements $w_{hk}$ of the matrix $\mathbf{W}$ that are defined equal to $q_{ij}^{\alpha \beta}$ when $q_{ij}^{\alpha \beta}>0$ or $0$ otherwise.
Notice that, given the element $w_{hk}$ of the matrix $\mathbf{W}$, it is always possible to obtain the specific weight $w_{ij}^{\alpha \beta}$, for each pair of sectors and economies, by reverting relation (\ref{eq:hk}).

  \subsection{Assessing the relevance of sectors and economies in terms of embodied energy flows}
  \label{MD-HITS_section}
  
  Now, we present centrality measures aimed at identifying economies and sectors that play a key role acting as collectors and distributors in the global energy flow system. A suitable measure for this aim is represented by hub and authority scores (see \cite{Kleinberg1999}, \cite{Kleinberg1999b}). In particular, we make
  use of an extension of the Kleinberg algorithm (\cite{Kleinberg1999}), called Multi-Dimensional HITS (MD-HITS), developed by \cite{Arrigo} and we adapt it to our framework. In this generalization for multilayer networks, not only nodes are ranked but also layers receive a score, ranking their capacity of linking important layers or being linked by hubs layers. Hence, the model defines four centrality vectors: two for nodes (hub and authority scores) and two for layers (broadcasting and receiving scores).
  It is noteworthy to say that node receives a high hub score if it belongs to a high broadcasting layer and has many links towards high authority nodes in layers with high receiving capabilities. Conversely, high authority would be awarded to nodes in high receiving layers with high hub nodes from high broadcasting layers pointing to them.\\
  To this aim, we consider two vectors of centrality for nodes: a  vector $\mathbf{x}=\left[x_i\right]$, with $i=1, \dots, N$, of hub scores and a vector $\mathbf{y}=\left[y_j\right]$, with $j=1, \dots, N$ of authority scores.
  Similarly, we have two vectors, $\mathbf{b}=\left[b_\alpha\right]$ and $\mathbf{z}=\left[z_\beta\right]$, with $\alpha,\beta=1, \dots, L$,  that account for the broadcasting and receiving capability of layers, respectively.\\
  Centrality scores are affected by a vector $\boldsymbol{\gamma}=\left[ \gamma_k\right]$, with $k=1, \dots, 4$. Each element of $\boldsymbol{\gamma}$ is such that $\gamma_k \in (0,1)$ and the following constraint must be (o is) satisfied $\sum_{k=1}^{4}\gamma_k=1$. Values of $\gamma_k$ allow to introduce a degree of flexibility in the model, highlighting the relevance of a specific score. Indeed, the more important a score, the higher the value of the corresponding weight.
  If no information is available concerning the importance of scores, a uniform choice can be inferred. In this case, we set $\gamma_k=0.25$, so that all the scores have the same importance in the model. \\
  The centrality scores are obtained via the multilayer generalization of the HITS algorithm by the following relations:
  \begin{align*}
  	x_{i} = \sum_{j,\alpha,\beta} \left( w_{ij}^{\alpha \beta}  y_{j}  b_{\alpha}  z_{\beta}\right) ^{\gamma_1}, \\
 	y_{j} = \sum_{i,\alpha,\beta} \left( w_{ij}^{\alpha \beta}  x_{i}  b_{\alpha}  z_{\beta}\right) ^{\gamma_2}, \\
 		b_{\alpha} = \sum_{i,j,\beta} \left( w_{ij}^{\alpha \beta}  x_{i}  y_{j} z_{\beta}\right) ^{\gamma_3}, \\
 			z_{\beta} = \sum_{i,j,\alpha} \left( w_{ij}^{\alpha \beta}  x_{i}  y_{j}  b_{\alpha}\right) ^{\gamma_4}.
  \end{align*}

As shown in \cite{Arrigo}, constraints on $\gamma$ assure that MD-HITS centrality exists. The centralities of nodes and layers can be obtained by solving the mutually-reinforcing recursive relationships defined above. In \cite{Arrigo}, the existence, uniqueness and maximality of the measure is given as well as a converging and fast parallel algorithm is provided. At each step all the vectors are non-negative and normalized such that their largest entry is equal to one.
	
\section{Numerical application}
\label{sec:na}

We provide here an application of the proposed methodology to a multilayer network based on the embodied energy flows among $26$ sectors/industries
and $189$ economies/countries.
Sectors and countries are listed in Appendix \ref{sectorscountrieslists} in Tables \ref{listsectors}, \ref{tab:tableofcountries_1} and \ref{tab:tableofcountries_2}, respectively. Data are extracted from EORA Global MRIO\footnote{\url{https://worldmrio.com/}} \cite{lenzen2013} for $2016$ since it provides the latest available information on the global network that can be downloaded freely from the EORA Global MRIO database. \\
The multilayer network is characterized by $189$ layers, representing the economies, and $26$ nodes, representing the sectors. Nodes are connected by weighted arcs describing the embodied energy flows specified in formula \eqref{qij_network}. Based on \cite{Chen2}, to calibrate the weights, we have considered six final-demand coupled system provided by EORA Global MRIO. For the purpose of applying formula \eqref{qij_network}, the total energy consumes are taken into account. Additionally, to extend the analysis on original input-output analysis, we include many sources of energy: coal, natural gas, petroleum, nuclear electricity, biomass and waste electricity, hydroelectricity and other renewable (geothermal, solar, tide, wave and wind) supplied by EORA Global MRIO. \\
The first part of the section is devoted to analyse the original input-output relations among sectors from different countries. Based on the information mined by $\mathbf{W}$, we summarise the main properties of industries and countries in the original input-output table. 
Then, we analyse the directed strengths at node and layer level to identify most relevant sectors and countries in terms of energy flows volume. The rest of this section is dedicated to the analysis of the MD-HITS approach results. In doing so, we investigate different directions of energy flows in the multilayer architecture and we detect key players in the system.

\subsection{Consumption}
\label{consumption_section}

\begin{figure}[htbp]
	\begin{subfigure}[t]{0.4\textwidth}
		\includegraphics[scale=0.25]{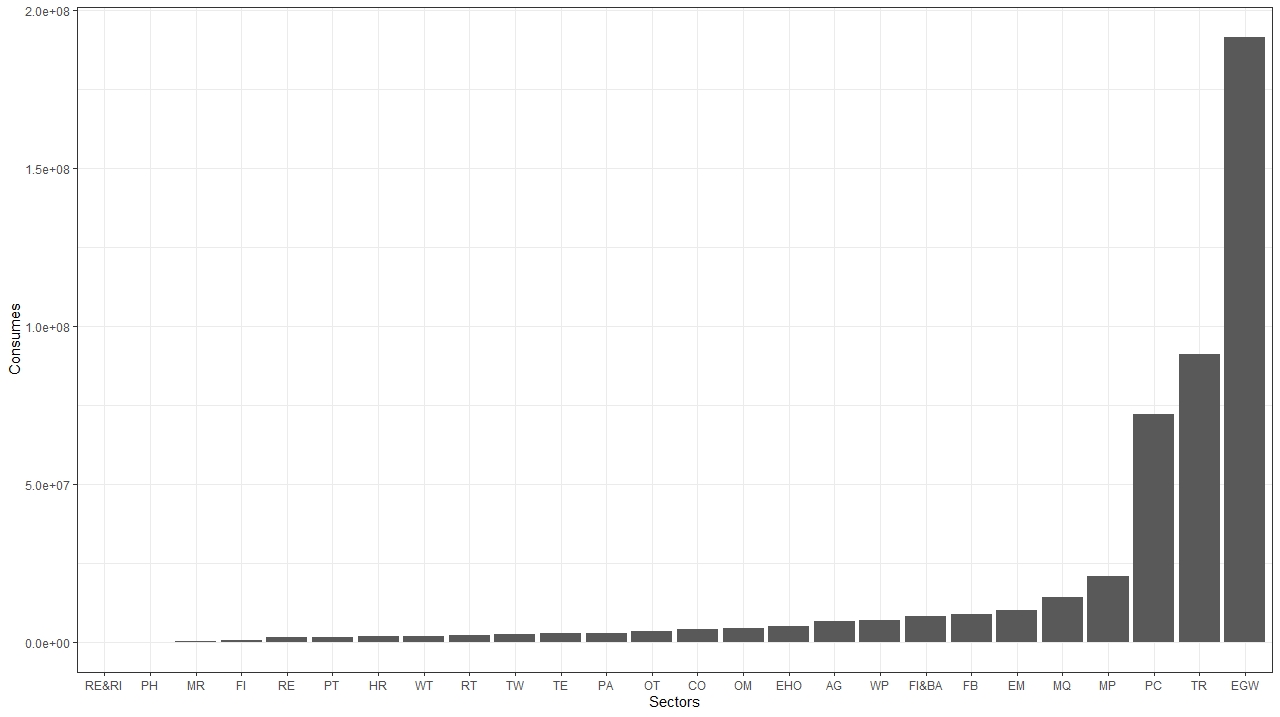}
		\caption{Sectors' consumption.}
		\label{cons_sector}
	\end{subfigure} \\
	\begin{subfigure}[t]{0.53\textwidth}
		\includegraphics[scale=0.25]{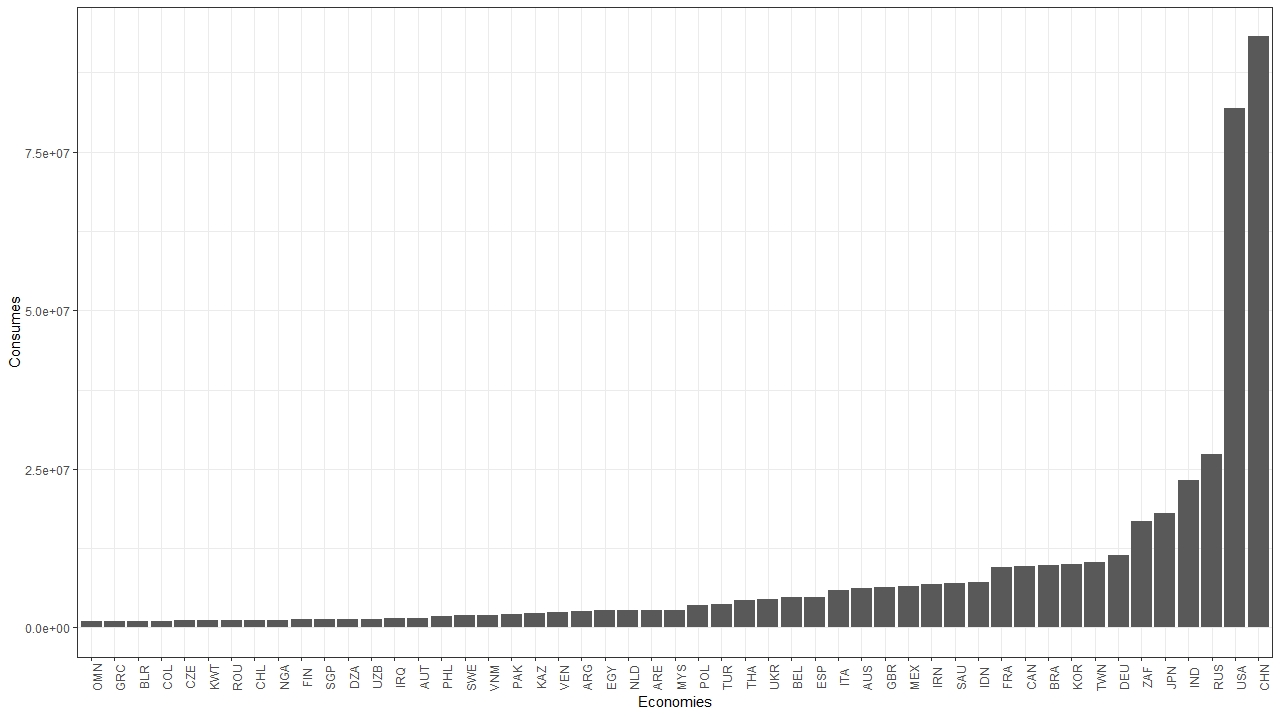}
		\caption{Top 50 economies' consumptions.}
		\label{cons_country}
	\end{subfigure} 
	\caption{Energy consumption of sectors and economies}
	\label{cons_tot}
\end{figure}
\begin{figure}
	\includegraphics[scale=0.3]{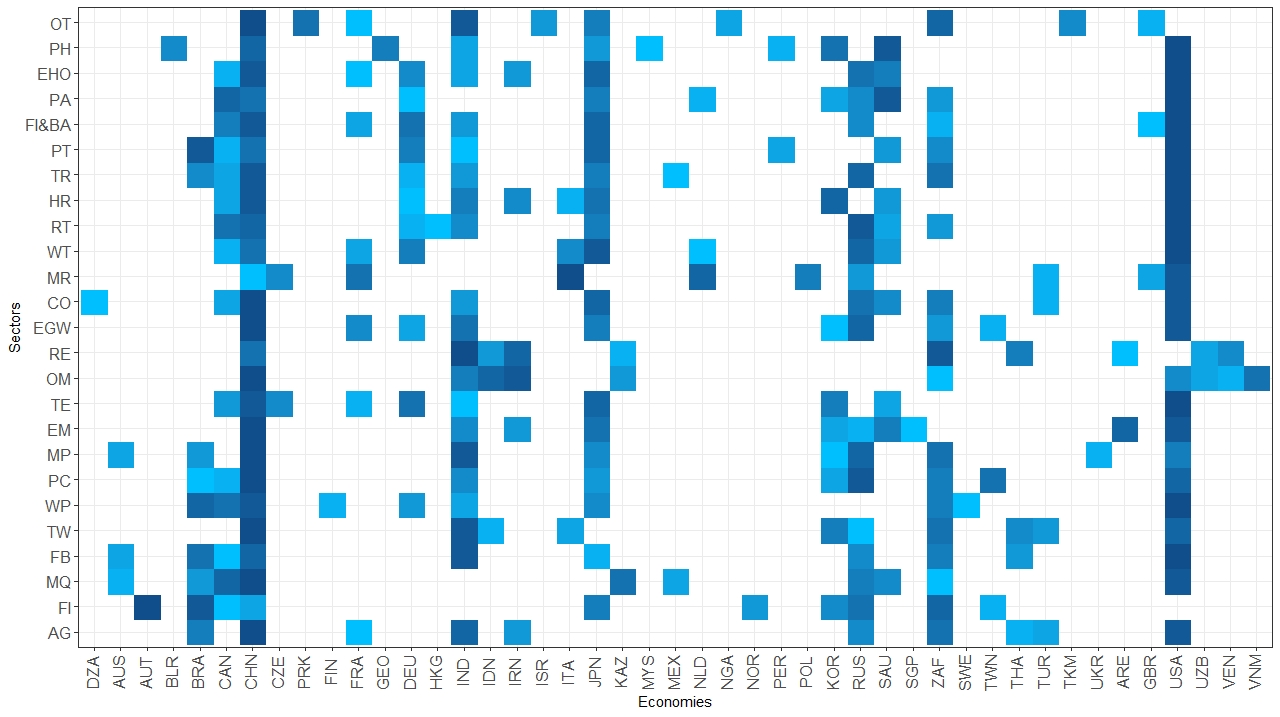}
	\caption{Top 10 countries in terms of energy consumption for each sector. Darker blue means higher consumes.}
	\label{important_edges}
\end{figure}

In order to analyse further results and the structure of the multilayer network, we start focusing on the energy consumption, reporting in Figure \ref{cons_tot} data split for each sector and country. \\
Focusing on sectors, Figure \ref{cons_sector} shows that the main sectors with the highest energy consumption are Electricity, Gas and Water (EGW), Petroleum, Chemical and non-Metallic Mineral Products (PC), Transport (TR) and Metal Products (MP). 
We notice that the above-mentioned sectors are characterised by elaborate end-products requiring several processing steps and inputs (e.g. iron or steel for Metal Product sector or fertilisers for the Petroleum, Chemical and non-Metallic Mineral Products sectors). We highlight that, contrary to what one might assume, Electricity, Gas and Water sector is in a prominent position. This result can be explained noticing that, in line with the analysis in \cite{WU2016}, the electricity generation consumes two mainly types of energy: the fuel (natural gas or coal) input energy and the energy used in fuel production, fuel transportation, materials, equipment and services. \\
Figure \ref{cons_country} shows the energy consumption of countries included in the analysis. In this case, the main consumers are China, United States, Russia, India and Japan. These results are in line, for instance, with the data about global electric energy consumption published  by the Energy Information Administration\footnote{\url{https://www.eia.gov/}}, Ernadata\footnote{\url{https://yearbook.enerdata.net/total-energy/world-consumption-statistics.html}} and the results in \cite{Chen2}.\\
Figure \ref{important_edges} displays for each sector the top ten countries in terms of energy consumption. This allows to catch the distribution of relevant economies in each sector. In particular, it is evident the prominent role of China and United States in all the sectors. Moreover, we observe that, considering the 26 sectors, China is indeed the biggest energy consumer in 10 of them and belongs to the top 5 in 23 sectors. The same kind of results can be read in \cite{shi2017}. Furthermore, we notice that 28 out of 30 arcs with higher weights are intra-layer connections in the layer that describes China economy. \\
United States is instead the top consumer in 12 sectors and in the top 5 in 21 sectors.\\
It is rather interesting to note how other relevant players, as Russia, Japan, India, South Africa, appear mainly exposed to sectors characterized by significant consumptions. \\
Looking at the European countries, we have that they are characterized by higher consumes in specific sectors related to the specificity of their economies. For instance, Austria is the biggest energy consumer in Fishing (FI) while Italy in Maintenance and Repair (MR). Because of its morphological structure and its lack of access to the sea, Austria depends largely on imports of marine and freshwater fish. The high energy consumption in fishing sector is explained noticing that, on the one hand, the fish demand is increased in the last years (see, \cite{acquaculture}) and, on the other hand, the fish market supply chain requires energy costs such as fuel input, transportation, materials, equipment and services, see \cite{huisingh2015}. For Italy, the result can be partially explained by the relevance of Maintenance and Repair industries in particular sectors (e.g. automotive).

\subsection{Analysis of Input-Output matrix}
\label{strengh_analisi}

\begin{figure}[htb]
	\begin{subfigure}[t]{0.51\textwidth}
		\includegraphics[scale=0.5]{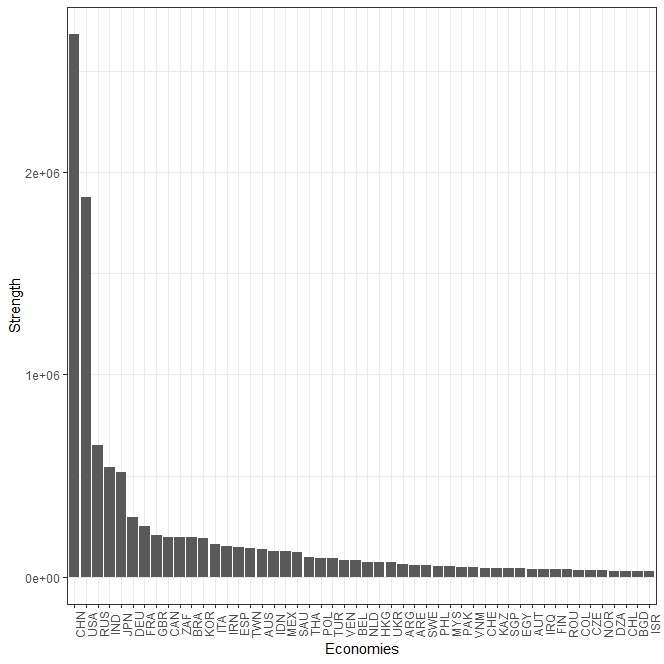}
		\caption{Economies' in-strength.}
		\label{country_in}
	\end{subfigure}\\
	\begin{subfigure}[t]{0.53\textwidth}
		\includegraphics[scale=0.5]{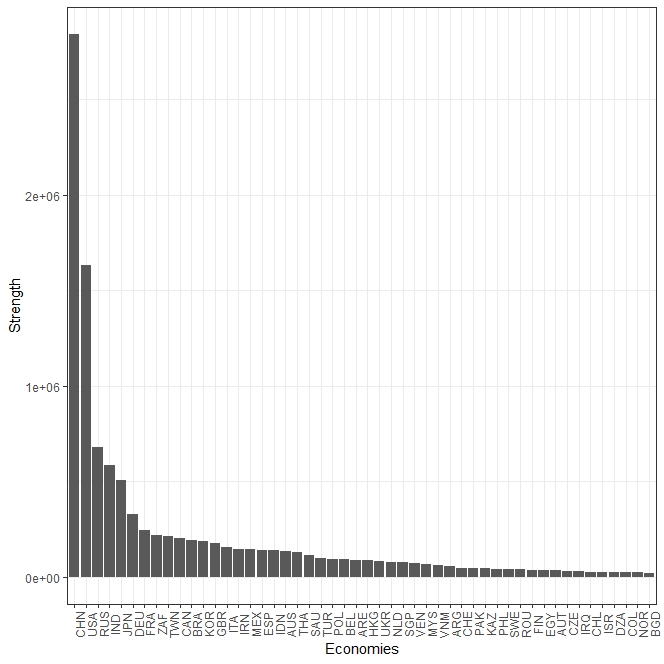}
		\caption{Economies' out-strength.}
		\label{country_out}
	\end{subfigure}
	\caption{Economies' in- and out-strengths.}
	\label{strength_paesi}
\end{figure}

\begin{figure}
	\begin{subfigure}[b]{0.51\linewidth}
		\includegraphics[scale=0.51]{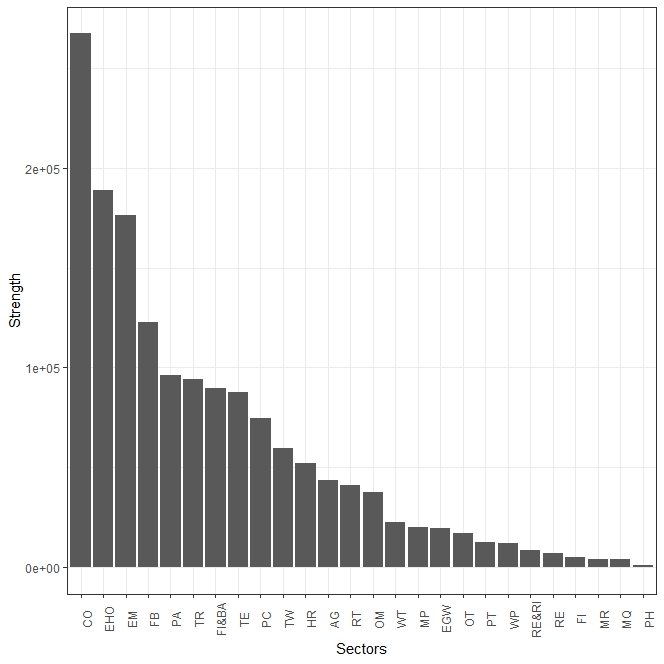}
		\caption{Sectors' in-strength.}
		\label{sect_int}
	\end{subfigure} \\
	\begin{subfigure}[b]{0.53\linewidth}
		\includegraphics[scale=0.3]{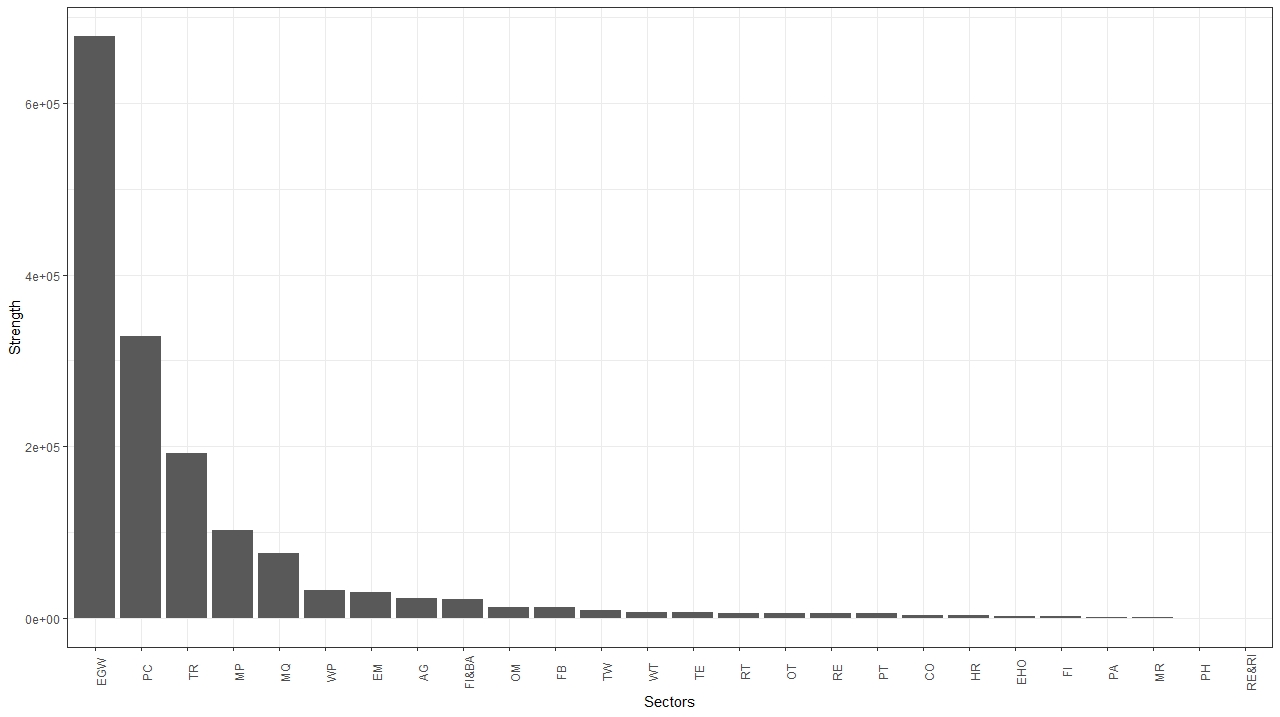}
		\caption{Sectors' out-strength.}
		\label{sect_out}
	\end{subfigure}
	\caption{Sectors' in- and out-strengths.}
	\label{Strength_settori}
\end{figure}

Figure \ref{strength_paesi} and \ref{Strength_settori} display the in- and out- strengths of each sector and economy considered in the multilayer network. Focusing on countries, in line with the literature (see, for example, \cite{Chen2}), we observe that the in- and out-strengths' rankings do not significantly change. At the first places, we find these countries characterized by a high internal consumption (see Figure \ref{cons_country}) and open to a higher quantities of international trades, i.e. China, United States, Russia, Japan. The results confirm not only the central positions of these countries in the commercial trade but also in the environmental damages. In fact, international trade consists mainly of tangible goods requiring fossil fuel-powered proceedings to be produced and exported (see \cite{huisingh2015}).
We have indeed that, although several kind of energies are taken into account in the consumes, these countries have relevant volumes of non-renewable energy consumes\footnote{Based on data supplied by \cite{lenzen2013}, in $2016$ renewable sources cover around $33\%$ of the total energy consumption worldwide.} (see, for example, \cite{owidenergy}). \\
Figure \ref{strength_paesi} confirms also that China is the most important contributors to the embodied energy flows (and therefore of pollution) distancing itself from the second largest country, the United States, by almost a million terajoule (TJ). Since the proposed approach looks to the entire supply chain, this result suggests that China plays also a central role as intermediate node for the passage of goods and services. \\
We highlight that Figures \ref{country_in} and \ref{country_out} show a dichotomous structure made by two main subgroups. A small subgroup of countries characterised by a high in- and out-strengths (i.e. an high embodied energy flow) and a large subgroup of nodes characterized by low in- and out-strength of embodied energy flows. We notice that countries belonging to the first subgroup are those characterised by a solid economy with an openness to the international scene. Moreover, we observe that the countries of the first group are those that either largely demand for goods (e.g. raw materials) from developing countries or are characterised by abundant natural resources (for example, natural gas for the Russia) which leads to a higher export volumes. The second subgroup includes countries that are smaller in terms of Gross Domestic Product (GDP) and, therefore, of international trade. 
Their economies are export-based and their low GDP reflects a low domestic demand for elaborate goods. This distinction highlights countries heterogeneity
underlying the different role in the environmental damages played by each country at the global level. \\
In Figures \ref{sect_int} and \ref{sect_out}, we observe significant differences in sectors' rankings based on in- and out-strength, respectively.  
The result is not surprising as each sector differs from the others for input factors (human capital, money, etc.) and processes to obtain its product. This complexity implies a high energy demand and it is therefore reasonable to expect that sectors producing intermediate or final goods are characterised by high in-strength. This hypothesis can be explained noticing that final products are placed at the end of the supply chain. Conversely, sectors selling primary or secondary goods will be characterised by high out-strength. In this case, those products are instead at the bottom of the supply chain of final goods.\\
In particular, the out-strength ranking shows three dominant sectors: Electricity, Gas and Water (EGW), Petroleum, Chemical and Non-Metallic Mineral Products (PC) and Transport (TR). These sectors distribute the highest energy flows to the other sectors. As stated above, they are involved in the production of primary/secondary goods and services. The strong out-strength (i.e. the powerful flow of energy) and the dependence on fossil fuels in both generation and processing, (see \cite{WANG2019}), identify these sectors among the most polluting (see, e.g., \cite{Birol2017}, \cite{canadell2007}, \cite{friedlingstein2010}, \cite{huisingh2015}). \\
In-strength ranking in Figure \ref{sect_int} presents three main sectors: Construction (CO), Education Health and Other Services (EHO) and Electrical and Machinery (EM). We notice that these sectors belong to the tertiary sector. Their outputs are placed at the end of the supply chain and therefore, these sectors incorporate the energy flows of all their miscellaneous input factors (as human capital, electricity, etc.). These results obtained here on a large scale are in line with findings found in \cite{GUO2021} for the distinct energy-use patterns of two China's megacities. The authors focus indeed on evolutionary features of the two cities' energy-use structures analysed from a consumption-based perspective. To summarise, for economies we observe on average a high correlation between in- and out-strength. Rank correlation between in- and out-strength is indeed higher than $0.9$. Vice versa, for sectors the rank correlation is around $0.2$. This is due to the fact that sectors that are relevant in terms of out-flows, not necessarily are also significant in terms of in-flows. The reason lies in the high diversity of arcs' weights among sectors due to a different amount of embodied energy needed for good production. These results emphasise while a monoplex analysis based only on economies, as traditionally employed in the  literature (see, \cite{Chen2} and \cite{qier}) could lead to lose meaningful insights.

\subsection{Results}\label{sec:r}

\begin{table}[ht]
	\begin{minipage}{0.3\textwidth}
		\begin{tabular}{cc}
			\toprule
			
				\hline
				\textbf{Hubs} & \textbf{Sector} \\ \hline \hline
				1 & EGW \\ 
				0.813 & PC \\ 
				0.672 & TR \\ 
				0.651 & MP \\
				0.560 & MQ \\ 
				0.451 & WP \\ 
				0.423 & EM \\ 
				0.398 & FI\&BA \\ 
				0.395 & AG \\ 
				0.330 & FB \\ 
				0.321 & OM \\ 
				0.294 & TW \\
				0.291 & RT \\ 
				0.287 & TE \\ 
				0.285 & WT \\ 
				0.271 & PT \\ 
				0.251 & RE \\ 
				0.239 & HR \\ 
				0.239 & OT \\ 
				0.228 & EHO \\ 
				0.224 & CO \\ 
				0.179 & FI \\ 
				0.148 & PA \\ 
				0.146 & MR \\ 
				0.085 & PH \\ 
				0 & RE\&RI \\ 
				\bottomrule
			\end{tabular}
		\end{minipage}
		\hfill
		\begin{minipage}{0.3\textwidth}
			\begin{tabular}{cc}
				\toprule
				
					\hline
					\textbf{Authority} & \textbf{Sector} \\ \hline \hline
					1 & CO \\ 
					0.870 & EHO \\ 
					0.865 & EM \\ 
					0.720 & FB \\ 
					0.693 & TE \\ 
					0.693 & FI\&BA \\ 
					0.688 & PA \\ 
					0.663 & TR \\ 
					0.640 & PC \\ 
					0.617 & TW \\ 
					0.608 & AG \\ 
					0.582 & HR \\ 
					0.538 & RT \\ 
					0.534 & OT \\ 
					0.529 & OM \\ 
					0.487 & MP \\ 
					0.449 & WT \\ 
					0.425 & WP \\ 
					0.424 & EGW \\
					0.414 & PT \\ 
					0.353 & MQ \\ 
					0.350 & FI \\ 
					0.299 & RE\&RI \\ 
					0.288 & RE \\ 
					0.261 & MR \\ 
					0.218 & PH \\ 
					\bottomrule
			\end{tabular}
		\end{minipage}
		
		\caption{Rankings of sectors based on hub and authority scores.}
		\label{resultsectors}
	\end{table}
	
	We present the results of the application of the MD-HITS described in Section \ref{MD-HITS_section} to the embodied energy flows multilayer network. We remind that with the proposed methodology, nodes and layers receive a score, ranking their capacity of linking or being linked by important nodes or layers. \\
	We start analysing the hub and authority rankings of the multilayer. First of all, we point out that the hub and authority rankings reflect the strength rankings shown and analysed in Section \ref{strengh_analisi}. This is quite standard in the literature since strength and eigenvector centralities are strongly correlated (see \cite{Estrada2012}). We go beyond the information given by the strength measure by analysing, through hubs and authorities, which are the most powerful sectors in the embodied energy flows network. We recall that a node is called hub if the node points to important nodes, which, in turn, are called authorities. \\
	Within this specification, as displayed in the left part of Table \ref{resultsectors}, we observe that the most important hubs are the Electricity (EGW), the Petroleum, Chemical and Non-Metallic Mineral Products (PC), the Transport (PT) and the Metal Products (MP) sectors.  
	At global level those sectors are the major contributors to the global economic development (see \cite{EUenergy}). From this perspective, the trade-off between environmental damage and economic growth rises up since the above-mentioned sectors are the most polluting ones as already pointed out in Section \ref{strengh_analisi}. In particular, it has been estimated that the energy sector and manufacturing sectors produce approximately the $42\%$ of CO$_2$ emission worldwide (see, e.g., \cite{friedlingstein2010} and \cite{huisingh2015}). \\
	The information provided by hubs is an important issue showing the fundamental pillars on which each policy-maker could build an environmental/economic strategy. Such a policy should simultaneously reduce industries environmental impact and maintain stable economic growth. In particular, the literature on environmental economics has analysed both qualitatively and quantitatively the alternative strategies to reduce the impact of these sectors on the environment (see, for example, \cite{Zhao2020} and \cite{MADURAIELAVARASAN2022112204}). \\
	Nowadays, a crucial element for each economy and for the society is indeed represented by the development of suitable energy management strategies. On the one hand, the management plan of the Directorate-General for Energy of the European Commission sets out the main objectives needed in order to reach energy efficiency in line with the European Green Deal\footnote{\url{https://ec.europa.eu/info/strategy/priorities-2019-2024/european-green-deal_en}}. On the other hand, the strict relationships between geopolitics and energy can affect the affordability of energy sources and the strategies that can be put in action (see, e.g., \cite{Christou2021}). The picture is further complicated by the fact that the identified hubs sectors produce primary and/or secondary goods. Without an adequate energy security strategy that ensures a stable supply of raw material (i.e., coal or natural gas for the electricity generation), the risk of a disruption in the supply chain is likely (see, e.g. \cite{OLATOMIWA2021}). Therefore, the action of policymakers should take into consideration two different aspects. On the one hand, an environmental strategy can be pursued to reduce the impact of these sectors on the environment. On the other hand, this strategy should assure that each country receives continuous and convenient energy flows (see \cite{su11154147} and  \cite{en14175268}). These two objectives are intrinsically linked. In order to reduce the environmental impact of energy generation, countries around the world are promoting the use of renewable energy sources. However, the energy generation from renewable sources is uncertain in terms of variability and intermittency as it is predominantly dependent on weather conditions. This carries the risk that the energy supply will not meet the demand. To avoid such a risk, fossil-fuel sources are used. Therefore, the availability and the trade of fossil-fuel raw material (i.e., natural gas or coal) are at the centre of geopolitical relations, which the policymaker has to look at carefully. 
	\\
	Within such a complex framework, the view provided only by hubs is limited since this centrality identifies the key sectors distributing energy flows. The analysis for a sustainable strategy is completed by an additional point of view provided by the study of authorities, which, in turn, identify the main sectors requiring energy flows to manufacture products. \\
	In the right column of Table \ref{resultsectors}, we provide the authority scores for each sector, i.e. nodes that require high volumes of energy flows. It is noteworthy that top authorities are: Construction (CO), Education, Health and Other Services (EHO), Electrical and Machinery (EM). We observe that
	central nodes are in this case sectors that supply final or intangible goods and that mostly belong to the tertiary or quaternary sectors.
	In particular, energy used directly in Construction (CO) and Electrical and Machinery (EM) sectors includes large volume of diesel for machinery as well as electricity for powering buildings and tools, presenting many opportunities to save energy. Proper strategies for an energy-efficient design can be indeed provided (see, e.g. \cite{melgar2022}). \\
	Additionally, the methodology can be used to assess the effect of endogenous and/or exogenous shocks on all the other economic sectors. For instance, it could be interesting to measure how different sectors and countries can be affected by the introduction of environmental policies of a specific country on polluting sectors (e.g. raising taxes) or by the economic and political effects on energy prices (see \cite{bashir2021} and  \cite{ionescu2020} for a complete review about environmental taxes analysis).
	
	\begin{table}[ht]
		\begin{minipage}{0.3\textwidth}
			\begin{tabular}{cc}
				\toprule
				
					\hline
					\textbf{Broadcasting} & \textbf{Economy} \\ \hline \hline
					1 & CHN \\ 
					0.825 & USA \\ 
					0.611 & RUS \\ 
					0.565 & IND \\ 
					0.558 & JPN \\ 		
					0.457 & DEU \\
					0.433 & FRA \\ 
					0.405 & CAN \\ 
					0.401 & KOR \\ 
					0.397 & GBR \\ 
					0.391 & TWN \\ 
					0.389 & ZAF \\ 
					0.386 & BRA \\ 
					0.369 & ITA \\ 
					0.366 & AUS \\ 
					0.366 & ESP \\ 
					0.365 & MEX \\ 
					0.361 & IDN \\ 
					0.353 & IRN \\ 
					0.338 & SAU \\ 
					0.334 & HKG \\ 
					0.328 & THA \\ 
					0.324 & POL \\ 
					0.318 & TUR \\ 
					0.306 & BEL \\ 
					\bottomrule
				\end{tabular}
			\end{minipage}
			\hfill
			\begin{minipage}{0.3\textwidth}
				\begin{tabular}{cc}
					\toprule
					
						\hline
						\textbf{Receiving} & \textbf{Economy} \\ \hline \hline
						1 & CHN \\ 
						0.804 & USA \\
						0.610 & RUS \\ 
						0.566 & IND \\ 
						0.555 & JPN \\ 
						0.464 & DEU \\ 
						0.429 & TWN \\ 
						0.428 & FRA \\ 
						0.407 & CAN \\ 
						0.397 & KOR \\ 
						0.389 & ZAF \\ 
						0.381 & MEX \\ 
						0.380 & GBR \\ 
						0.378 & BRA \\ 
						0.368 & ITA \\ 
						0.363 & AUS \\ 
						0.360 & IDN \\ 
						0.357 & ESP \\ 
						0.350 & THA \\ 
						0.346 & IRN \\ 
						0.343 & HKG \\ 
						0.326 & SAU \\ 
						0.317 & POL \\ 
						0.315 & TUR \\ 
						0.313 & SGP \\ 
						\bottomrule
					\end{tabular}
				\end{minipage}
				\caption{Top 25 economies based on broadcasting and receiving scores.}
				\label{resultcountries}
			\end{table}
			
			\begin{figure}[htbp]
				\centering
				\begin{subfigure}[h]{0.45\textwidth}
					\includegraphics[width=\textwidth]{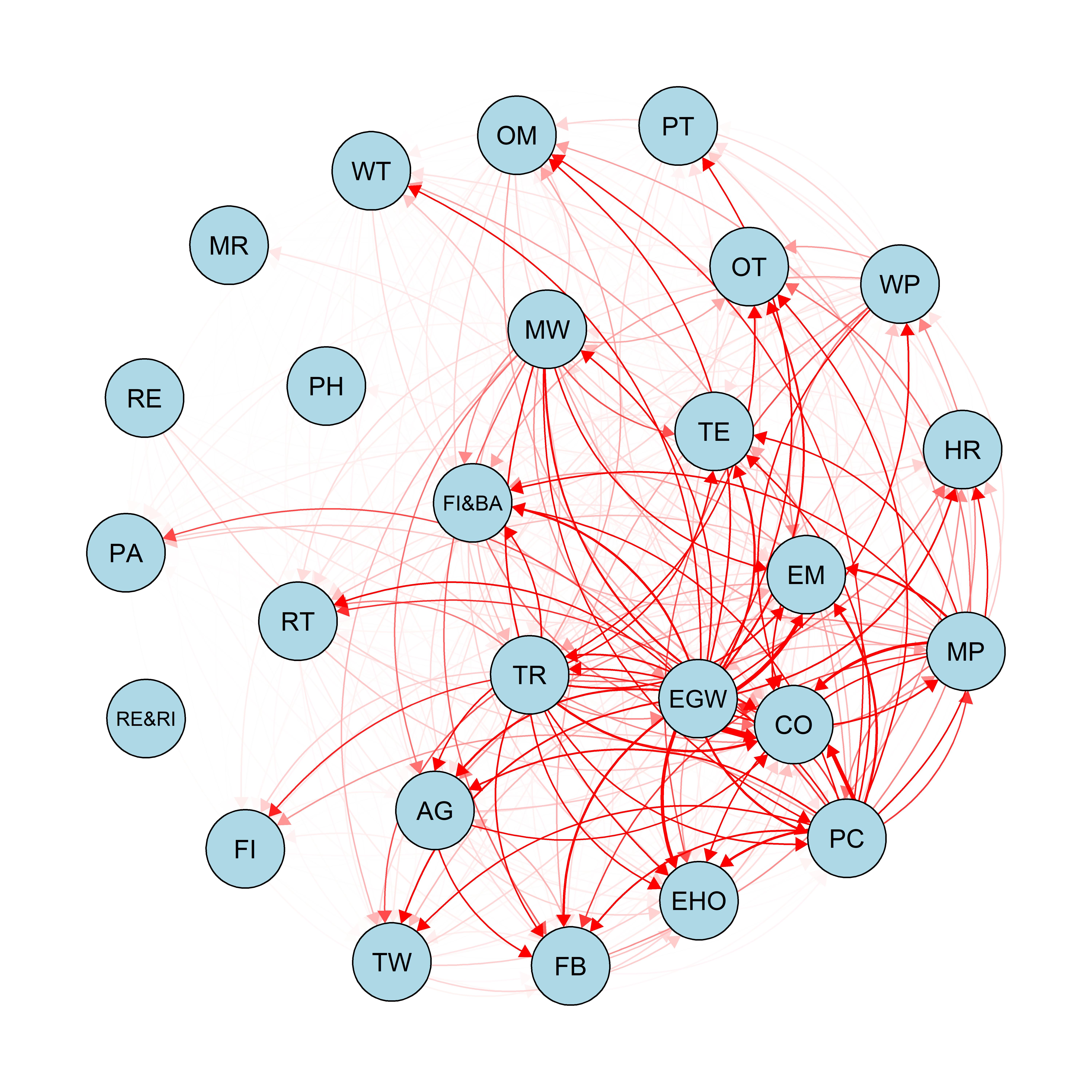}
					\caption{China network representation.}
					\label{China_graph}
				\end{subfigure}
				\begin{subfigure}[h]{0.45\textwidth}
					\includegraphics[width=\textwidth]{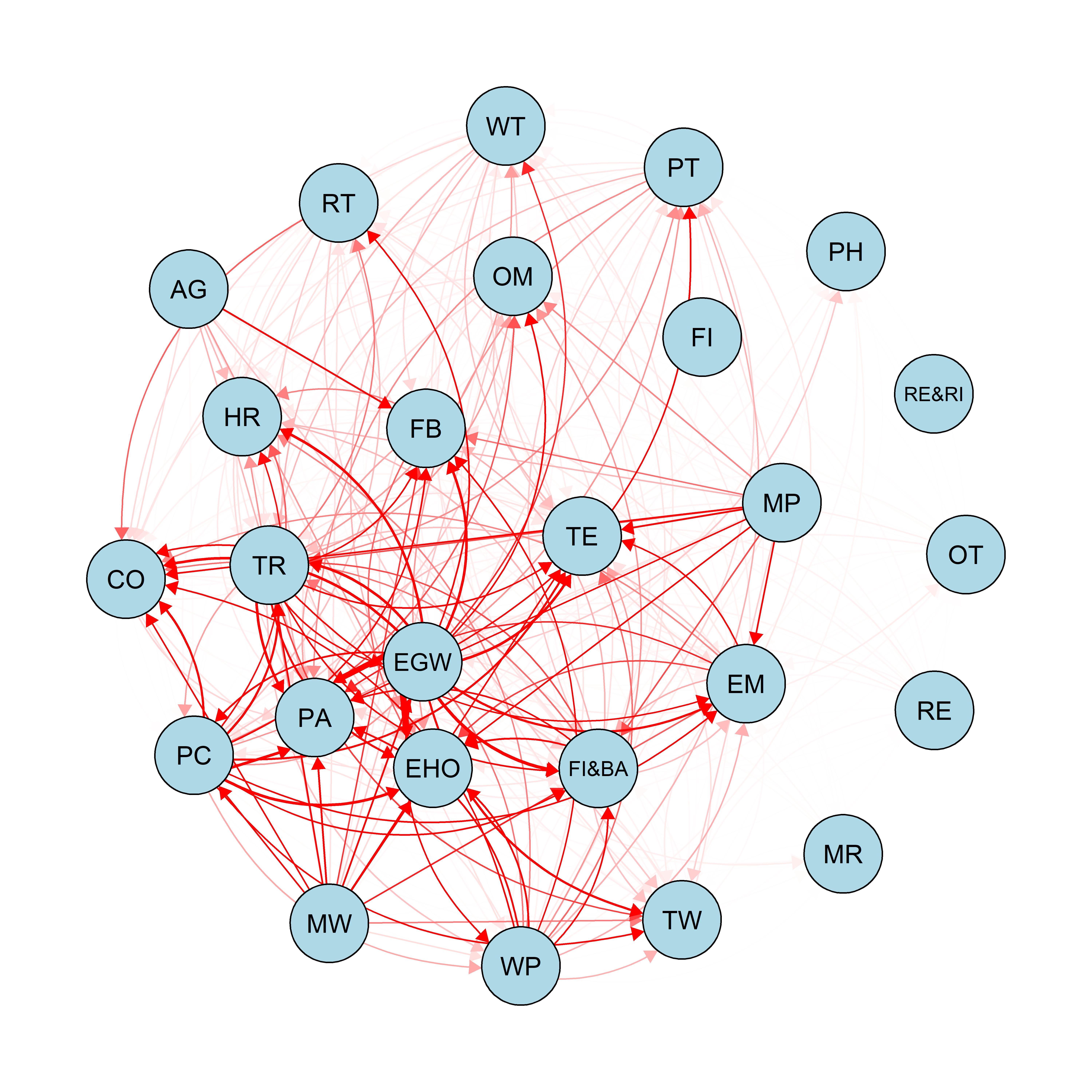}
					\caption{United States network representation.}
					\label{USA_graph}
				\end{subfigure}\\
				\begin{subfigure}[h]{0.45\textwidth}
					\includegraphics[width=\textwidth]{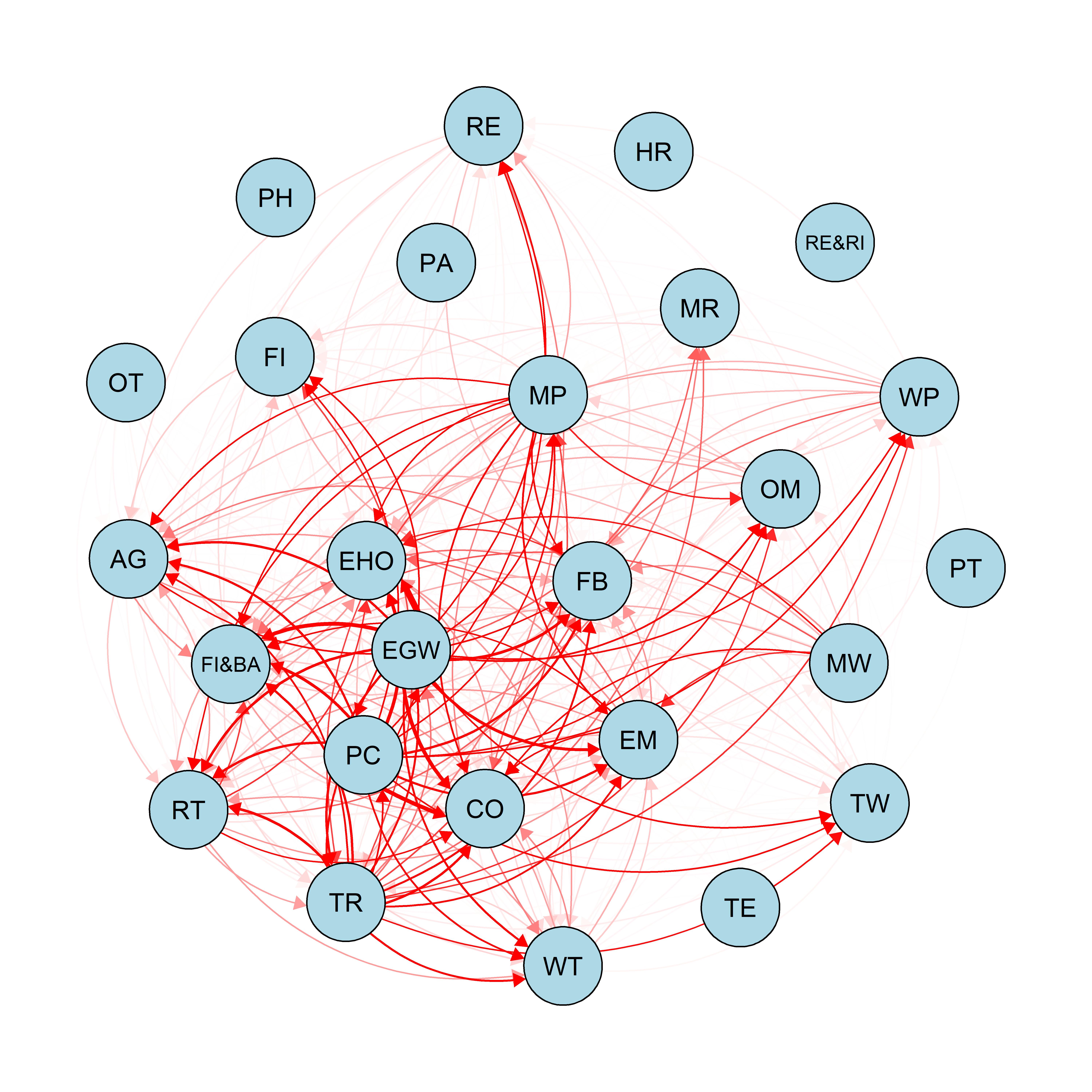}
					\caption{Russia network representation.}
					\label{Russia_graph}
				\end{subfigure}
				\caption{Monolayer network for three countries. Each network has been obtained considering only arcs between sectors of the country. The size of each arc is proportional to its weight.}
				\label{Networktopologies}
			\end{figure}
			
			In the same spirit of the definitions of hubs and authorities, we analyse the broadcasting and receiving centrality rankings for the layers, which, we remind, in this work represent the countries. \\
			When analysing the layers, broadcasting and receiving are the counterparts of hubs and authorities respectively. Table \ref{resultcountries} shows the broadcasting and receiving scores for the top 25 countries. We observe that, unlike the sectors, broadcasting and receiving rankings show no significant differences. China is still the most central and powerful nation in embodied energy flows, followed by the Russia, United States, Japan, India. The result is in line with the in- and out-strength analysis of Section \ref{strengh_analisi}. 
			Central countries are indeed also the top five countries based on total final consumption. Some of them trade less extensively than others, due to having large domestic markets and limited interconnections with neighbouring regions. According to the World Energy and Climate Statistics published by Enerdata\footnote{\url{https://yearbook.enerdata.net/total-energy/world-import-export-statistics.html}}, Eastern Asian countries (as China, India, Japan and South Korea) and several European countries (in particular Italy and Germany) are characterized by a significant negative balance trade, while Russia and Middle East (e.g., Saudi Arabia and Emirates) have indeed an opposite behaviour being important exporters. In this case, since the network considers embodied energy flows and it is affected by both energy consumes and volume of trades in the domestic and in the international markets, we observe that most central countries appear relevant both in terms of consumes and demands.\\
			 Additionally, we display in Figure \ref{Networktopologies} monolayer networks for China, Russia and United States, based on embodied energy flows between sectors of the same country. It is noticeable the high value of embodied energy flow exchanged internally by these countries. Furthermore, we notice that relevant sectors, based on multilayers hubs and scores and reported in Table \ref{resultsectors}, play a crucial role for these countries (see, for instance, node EGW in the three networks). \\
			Based on broadcasting and receiving rankings, it could be observed that the top-ranked countries are representative of three largest clusters detected by community detection methods on the international trade network (see, for example, \cite{bartesaghi2020}, \cite{Blochl2011}, \cite{DeBenedictis2011}, \cite{ercsey2012}, \cite{Fagiolo2010}, \cite{grassi2021}, \cite{Serrano2007} and \cite{tzekina2008}): the American cluster, the Pacific cluster and the European cluster. 
			In addition to the information mined by the mesoscale structure, the proposed approach highlights the pivotal countries of each community. This allows not only to stress the economic power of a country within the global scenario but also at cluster level. Furthermore, through the multilayer approach, it is possible to trace the intermediate production steps of each good. This makes it possible to highlight the role played by the mesoscale structure of the international trade in the good supply chain emphasizing the impact of the former on the price of the final good.

			\section{Conclusions}
			\label{sec:conc}
			In this paper, we extend \replaced{classical}{the} input-output analysis considering the environmental impact through a multilayer network. In particular, we construct a multilayer network \replaced{where sectors of each economy are connected by weighted arcs representing the directed embodied energy flows.}{connecting sectors and economies.} \deleted{Sector represents the nodes on each layer, which represents the economy/country. Nodes on layers are connected by directed and weighted arcs representing the embodied energy flows among them.} \deleted{In this work,} \replaced{A}{a} multilayer generalisation of hub and authority measures is applied to assign a score both to nodes and layers. \replaced{. In this way, we detect relevant sectors and economies in the whole system.\\}{\textit{}In doing so, we simultaneously rank sectors and economies identifying their capacity of linking important nodes/layers or being linked by hubs nodes/layers. \\} 
			\replaced{Notably, hubs are sectors devoted to produce primary and/or secondary goods. }{The results identify as hubs sectors the ones devoted to produce primary and/or secondary goods.} These sectors are \added{also} the major contributors to the global economic development and environmental degradation. Sectors, \replaced{that supply final or intangible goods and that mostly belong to the tertiary or quaternary sectors, have been instead identified as authorities.}{that have been identified as authorities are those that supply final or intangible goods and that mostly belong to the tertiary or quaternary sectors.} 
			\replaced{According to economies,}{Dealing with countries,} the rankings of broadcasting and receiving scores do not show significant \replaced{differences.}{changes.} \added{Relevant countries, in terms of consumes and demands, are the most important players in terms of embodied energy flows.}
			\deleted{Both rankings highlight that China, United States and Russia are the most important economies worldwide in terms of energy flows. The same kind of result has been also found by analysing the countries' strength measure confirming that the most central countries appear relevant both in terms of consumes and demands. \\}
			\added{It is worth to be mentioned that the proposed approach can provide useful insights for policymakers. Indeed, main results can represent a tool for the development of an environmental strategy based on a trade-off between reducing the environmental degradation and maintaining the economic growth. The multilayer structure, providing an overview of the energy supply chain, allows to trace the effects on the system of exogenous and endogenous shocks as, for instance, changes in countries’ environmental policies.\\ }
			\deleted{The problem analysed in this work can provide useful insights for policymakers. A first result concerns the needs for building an environmental strategy that reduces the impact of hubs sector on the environment. This fact highlights the need for balancing the trade-off between environmental degradation and economic growth. In particular, an approach based on multilayer network is useful to identify the sectors and foreign countries that could be affected by the economic repercussions of a country's environmental policies. \\}
			\added{	Further research on this topic will regard the analysis of alternative energy sources. Indeed, some renewable sources, as wind and solar power, can be characterized by a variability that could lead to a reduction in the energy supply. Therefore, there is the possibility that the gap between demand and supply of energy has to be filled by polluting fossil-fuel sources. On the other hand, the availability and the trade of fossil-fuel sources are at the centre of geopolitical relations, which the policymaker has to look at carefully.}
			\deleted{Moreover, the analysed multilayer network allows the policymaker to have an overview of the energy supply chain. This vision allows him/her to complete the framework for an energy policy based on the use of renewable resources ensuring a continuous and convenient energy flows. These two targets are strictly connected. Renewable sources are characterized by variability and intermittency that could lead to a reduction in the energy supply; the gap between demand and supply of energy has to be filled by polluting fossil-fuel sources. Finally, the availability and the trade of fossil-fuel sources are at the centre of geopolitical relations, which the policymaker has to look at carefully.}

			\appendix
			
			\section{List of sectors and countries}
			\label{sectorscountrieslists}
			
			\begin{table}
				\begin{tabular}{lc}
					Sectors	& Code \\
					\hline \hline
					Agriculture	& AG \\
					Fishing	& 	FI\\
					Mining and Quarrying	& 	MW\\
					Food and Beverages 	& 	FB\\
					Textiles and Wearing Apparel 	& 	TW\\
					Wood and Paper 	& 	WP\\
					Petroleum, Chemical and Non-Metallic Mineral Products	& 	PC\\
					Metal Products 	& 	MP\\
					Electrical and Machinery 	& 	EM\\
					Transport Equipment		& TE\\
					Other Manufacturing 	& 	OM\\
					Recycling 	& 	RE\\
					Electricity, Gas and Water	& 	EGW\\
					Construction	& 	CO\\
					Maintenance and Repair	& 	MR\\
					Wholesale Trade 	& 	WT\\
					Retail Trade	& 	RT\\
					Hotels and Restaurants		& HR\\
					Transport	& 	TR\\
					Post and Telecommunications	& 	PT\\
					Financial Intermediation and Business Activities 	& 	FI\&BA\\
					Public Administration 	& 	PA\\
					Education, Health and Other Services	& 	EHO\\
					Private Households 	& 	PH\\
					Others		& OT\\
					Re-export and Re-import 	& 	RE\&RI\\
					\hline
				\end{tabular}
				\caption{List of sectors.}
				\label{listsectors}
			\end{table}
			
			\begin{table}[ht]
				\begin{minipage}{0.3\textwidth}
					\begin{tabular}{lc}
						\toprule
						\textbf{Country} & \textbf{Code} \\
						\midrule
						Afghanistan & AFG \\
						Albania & ALB \\
						Algeria & DZA \\
						Andorra & AND \\
						Angola & AGO \\
						Antigua & ATG \\
						Argentina & ARG \\
						Armenia & ARM \\
						Aruba & ABW \\
						Australia & AUS \\
						Austria & AUT \\
						Azerbaijan & AZE \\
						Bahamas & BHS \\
						Bahrain & BHR \\
						Bangladesh & BGD \\
						Barbados & BRB \\
						Belarus & BLR \\
						Belgium & BEL \\
						Belize & BLZ \\
						Benin & BEN \\
						Bermuda & BMU \\
						Bhutan & BTN \\
						Bolivia & BOL \\
						Bosnia Herzegovina & BHI \\
						Botswana & BWA \\
						Brazil & BRA \\
						British Virgin Islands & VGB \\
						Brunei & BRN \\
						Bulgaria & BGR \\
						Burkina Faso & BFA \\
						Burundi & BDI \\
						Cambodia & KHM \\
						Cameroon & CMR \\
						Canada & CAN \\
						Cape Verde & CPV \\
						Cayman Islands & CYM \\
						Central African Republic & CAF \\
						Chad & TCD \\
						Chile & CHL \\
						China & CHN \\
						Colombia & COL \\
						Congo & COG \\
						Costa Rica & CRI \\
						Cote d'Ivoire & CIV \\
						Croatia & HRV \\
						Cuba & CUB \\
						Cyprus & CYP \\
						\bottomrule
					\end{tabular}
				\end{minipage}
				\hfill
				\begin{minipage}{0.3\textwidth}
					\begin{tabular}{lc}
						\toprule
						\textbf{Country} & \textbf{Code} \\
						\midrule
						Czech Republic & CZE \\
						Denmark & DNK \\
						Djibouti & DJI \\
						Dominican Republic & DOM \\
						Congo & COG \\
						Ecuador & ECU \\
						Egypt & EGY \\
						El Salvador & SLV \\
						Eritrea & ERI \\
						Estonia & EST \\
						Ethiopia & ETH \\
						Fiji & FJI \\
						Finland & FIN \\
						Former USSR & SUN \\
						France & FRA \\
						French Polynesia & PYF \\
						Gabon & GAB \\
						Gambia & GMB \\
						Gaza Strip & PSE \\
						Georgia & GEO \\
						Germany & DEU \\
						Ghana & GHA \\
						Greece & GRC \\
						Greenland & GRL \\
						Guatemala & GTM \\
						Guinea & GIN \\
						Guyana & GUY \\
						Haiti & HTI \\
						Honduras & HND \\
						Hong Kong & HKG \\
						Hungary & HUN \\
						Iceland & ISL \\
						India & IND \\
						Indonesia & IDN \\
						Iran & IRN \\
						Iraq & IRQ \\
						Ireland & IRL \\
						Israel & ISR \\
						Italy & ITA \\
						Jamaica & JAM \\
						Japan & JPN \\
						Jordan & JOR \\
						Kazakhstan & KAZ \\
						Kenya & KEN \\
						Kuwait & KWT \\
						Kyrgyzstan & KGZ \\
						\bottomrule
					\end{tabular}
				\end{minipage}
				\caption{List of countries - I.}
				\label{tab:tableofcountries_1}
			\end{table}
			
			\begin{table}[ht]
				\begin{minipage}{0.3\textwidth}
					\begin{tabular}{lc}		
						\toprule
						\textbf{Country} & \textbf{Code} \\
						\midrule
						Laos & LAO \\
						Latvia & LVA \\
						Lebanon & LBN \\
						Lesotho & LSO \\
						Liberia & LBR \\
						Libya & LBY \\
						Liechtenstein & LIE \\
						Lithuania & LTU \\
						Luxembourg & LUX \\
						Macao & MAC \\
						Madagascar & MDG \\
						Malawi & MWI \\
						Malaysia & MYS \\
						Maldives & MDV \\
						Mali & MLI \\
						Malta & MLT \\
						Mauritania & MRT \\
						Mauritius & MUS \\
						Mexico & MEX \\
						Moldova & MDA \\
						Monaco & MCO \\
						Mongolia & MNG \\
						Montenegro & MNE \\
						Morocco & MAR \\
						Mozambique & MOZ \\
						Myanmar & MMR \\
						Namibia & NAM \\
						Nepal & NPL \\
						Netherlands & NLD \\
						Netherlands Antilles & ANT \\
						New Caledonia & NCL \\
						New Zealand & NZL \\
						Nicaragua & NIC \\
						Niger & NER \\
						Nigeria & NGA \\
						North Korea & PRK \\
						Norway & NOR \\
						Oman & OMN \\
						Pakistan & PAK \\
						Panama & PAN \\
						Papua New Guinea & PNG \\
						Paraguay & PRY \\
						Peru & PER \\
						Philippines & PHL \\
						Poland & POL \\
						Portugal & PRT \\
						Qatar & QAT \\
						\bottomrule
					\end{tabular}
				\end{minipage}
				\hfill
				\begin{minipage}{0.3\textwidth}
					\begin{tabular}{lc}		
						\toprule
						\textbf{Country} & \textbf{Code} \\
						\midrule					
						
						Romania & ROU \\
						Russia & RUS \\
						Rwanda & RWA \\
						Samoa & WSM \\
						San Marino & SMR \\
						Sao Tome and Principe & STP \\
						Saudi Arabia & SAU \\
						Senegal & SEN \\
						Serbia & SRB \\
						Seychelles & SYC \\
						Sierra Leone & SLE \\
						Singapore & SGP \\
						Slovakia & SVK \\
						Slovenia & SVN \\
						Somalia & SOM \\
						South Africa & ZAF \\
						South Korea & KOR \\
						South Sudan & SSD \\
						Spain & ESP \\
						Sri Lanka & LKA \\
						Sudan & SDN \\
						Suriname & SUR \\
						Swaziland & SWZ \\
						Sweden & SWE \\
						Switzerland & CHE \\
						Syria & SYR \\
						Taiwan & TWN \\
						Tajikistan & TJK \\
						Tanzania & TZA \\
						TFYR Macedonia & MKD \\
						Thailand & THA \\
						Togo & TGO \\
						Trinidad and Tobago & TTO \\
						Tunisia & TUN \\
						Turkey & TUR \\
						Turkmenistan & TKM \\
						United Arab Emirates & ARE \\
						Uganda & UGA \\
						United Kingdom & GBR \\
						Ukraine & UKR \\
						Uruguay & URY \\
						United States & USA \\
						Uzbekistan & UZB \\
						Vanuatu & VUT \\
						Venezuela & VEN \\
						Vietnam & VNM \\
						Yemen & YEM \\
						Zambia & ZMB \\
						Zimbabwe & ZWE\\
						\bottomrule
					\end{tabular}
				\end{minipage}
				\caption{List of countries - II.}
				\label{tab:tableofcountries_2}
			\end{table}


\bibliographystyle{plain}
\bibliography{Myref}   

\begin{thebibliography}{10}

\bibitem{Arrigo}
F.~Arrigo and F.~Tudisco.
\newblock Multi-dimensional, multilayer, nonlinear and dynamic hits.
\newblock {\em Proceedings of the 2019 SIAM International Conference on Data
  Mining (SDM)}, pages 369--377, 2019.

\bibitem{bartesaghi2020}
P.~Bartesaghi, G.~P. Clemente, and R.~Grassi.
\newblock Community structure in the {W}orld {T}rade {N}etwork based on
  communicability distances.
\newblock {\em Journal of Economic Interaction and Coordination}, pages 1--37,
  2020.

\bibitem{bashir2021}
M.~F. Bashir, B.~Ma, B.~Komal, and M.~A. Bashir.
\newblock Analysis of environmental taxes publications: a bibliometric and
  systematic literature review.
\newblock {\em Environmental Science and Pollution Research},
  28(16):20700--20716, 2021.

\bibitem{Birol2017}
F.~Birol.
\newblock {CO}$_2$ emissions from fuel combustion - {H}ighlights 2017.
\newblock Technical report, International {E}nergy {A}gency, 2017.

\bibitem{acquaculture}
K.~Blaas.
\newblock Aquaculture 2020 - {A}ustrian strategy to increase the national fish
  production.
\newblock Technical report, Federal Ministry of {A}griculture, {F}orestry,
  {E}nvironment and {W}ater Management, Stubenring 1, A-1010 Vienna, 2021.

\bibitem{Blochl2011}
F.~Bl{\"o}chl, F.~J. Theis, F.~Vega-Redondo, and E.~O’N. Fisher.
\newblock Vertex centralities in input-output networks reveal the structure of
  modern economies.
\newblock {\em Physical Review E}, 83(4):046127, 2011.

\bibitem{Bonacichb}
P.~Bonacich.
\newblock Factoring and weighting approaches to status scores and clique
  identification.
\newblock {\em Journal of Mathematical Sociology}, 2:113--120, 1972.

\bibitem{Bonacicha}
P.~Bonacich.
\newblock A technique for analyzing overlapping memberships.
\newblock {\em Sociological Methodology}, pages 176--185, 1972.

\bibitem{Brown}
M.~T. Brown and R.~A. Herendeen.
\newblock Embodied energy analysis and emergy analysis: a comparative view.
\newblock {\em Ecological Economics}, 19(3):219--235, 1996.

\bibitem{canadell2007}
J.~G. Canadell, C.~Le~Qu{\'e}r{\'e}, M.~R. Raupach, C.~B. Field, E.~T.
  Buitenhuis, P.~Ciais, T.~J. Conway, N.~P. Gillett, R.~A. Houghton, and
  G.~Marland.
\newblock Contributions to accelerating atmospheric {CO}$_2$ growth from
  economic activity, carbon intensity, and efficiency of natural sinks.
\newblock {\em Proceedings of the national academy of sciences},
  104(47):18866--18870, 2007.

\bibitem{Chen2}
B.~Chen, J.S. Li, X.F. Wu, M.Y. Han, L.~Zeng, Z.~Li, and G.~Q. Chen.
\newblock Global energy flows embodied in international trade: A combination of
  environmentally extended input–output analysis and complex network
  analysis.
\newblock {\em Applied Energy}, 210:98--107, 2018.

\bibitem{EUenergy}
European {C}ommission and {D}irectorate-{G}eneral~for {E}nergy.
\newblock {\em {EU} energy in figures : statistical pocketbook 2018}.
\newblock Publications Office, 2018.

\bibitem{DeBenedictis2011}
L.~De~Benedictis and L.~Tajoli.
\newblock The world trade network.
\newblock {\em The World Economy}, 34(8):1417--1454, 2011.

\bibitem{Du}
R.~Du, Y.~Wang, G.~Dong, L~Tian, Y.~Liu, M.~Wang, and G.~Fang.
\newblock A complex network perspective on interrelations and evolution
  features of international oil trade.
\newblock {\em Applied Energy}, 196:2002--2013, 2016.

\bibitem{MADURAIELAVARASAN2022112204}
M.~R. Elavarasan, R.~Pugazhendhi, M.~Irfan, L.~Mihet-Popa, I.~A. Khan, and
  P.~E. Campana.
\newblock State-of-the-art sustainable approaches for deeper decarbonization in
  {E}urope – {A}n endowment to climate neutral vision.
\newblock {\em Renewable and Sustainable Energy Reviews}, 159:112204, 2022.

\bibitem{ercsey2012}
M.~Ercsey-Ravasz, Z.~Toroczkai, Z.~Lakner, and J.~Baranyi.
\newblock Complexity of the international agro-food trade network and its
  impact on food safety.
\newblock {\em PloS one}, 7(5):e37810, 2012.

\bibitem{Estrada2012}
E.~Estrada.
\newblock {\em The structure of complex networks: theory and applications}.
\newblock Oxford University Press, 2012.

\bibitem{Fagiolo2010}
G.~Fagiolo, J.~Reyes, and S.~Schiavo.
\newblock The evolution of the world trade web: a weighted-network analysis.
\newblock {\em Journal of Evolutionary Economics}, 20(4):479--514, 2010.

\bibitem{friedlingstein2010}
P.~Friedlingstein, R.~A. Houghton, G.~Marland, J.~Hackler, T.~A. Boden, T.~J.
  Conway, J.~G. Canadell, M.~R. Raupach, P.~Ciais, and C.~Le~Qu{\'e}r{\'e}.
\newblock Update on {CO}$_2$ emissions.
\newblock {\em Nature geoscience}, 3(12):811--812, 2010.

\bibitem{grassi2021}
R.~Grassi, P.~Bartesaghi, S.~Benati, and G.~P. Clemente.
\newblock Multi-attribute community detection in international trade network.
\newblock {\em Networks and Spatial Economics}, 21(3):707--733, 2021.

\bibitem{GUO2021}
S.~Guo, Y.~Li, Ping He, Haosong Chen, and Jing Meng.
\newblock Embodied energy use of {C}hina's megacities: A comparative study of
  {B}eijing and {S}hanghai.
\newblock {\em Energy Policy}, 155:112243, 2021.

\bibitem{Hao}
X.~Hao, H.~An, H.~Qi, and Gao X.
\newblock Evolution of the exergy flow network embodied in the global fossil
  energy trade: based on complex network.
\newblock {\em Applied Energy}, 196:1515--1522, 2016.

\bibitem{Harary}
F.~Harary.
\newblock {\em Graph theory.}
\newblock Addison-Wesley, Reading, MA, 1969.

\bibitem{huisingh2015}
D.~Huisingh, Z.~Zhang, J.~C. Moore, Q.~Qiao, and Q.~Li.
\newblock Recent advances in carbon emissions reduction: policies,
  technologies, monitoring, assessment and modeling.
\newblock {\em Journal of Cleaner Production}, 103:1--12, 2015.

\bibitem{ionescu2020}
L.~Ionescu.
\newblock The economics of the carbon tax: Environmental performance,
  sustainable energy, and green financial behavior.
\newblock {\em Geopolitics, History, and International Relations},
  12(1):101--107, 2020.

\bibitem{Kites}
J.~Kites.
\newblock An introduction to environmentally-extended input-output analysis.
\newblock {\em Resources}, 11:489--503, 2013.

\bibitem{Kivela}
M.~Kivel{\"a}, A.~Arenas, M.~Barthelemy, J.~P. Gleeson, Y.~Moreno, and M.~A.
  Porter.
\newblock Multilayer networks.
\newblock {\em Journal of Complex Networks,}, 2:203--271, 2014.

\bibitem{Kleinberg1999}
J.~M. Kleinberg.
\newblock Authoritative sources in a hyperlinked environment.
\newblock {\em Journal of the ACM (JACM)}, 46(5):604--632, 1999.

\bibitem{Kleinberg1999b}
J.~M. Kleinberg.
\newblock Hubs, authorities, and communities.
\newblock {\em Journal of the ACM (JACM)}, 31(4):5--es, 1999.

\bibitem{lenzen2013}
M.~Lenzen, D.~Moran, K.~Kanemoto, and A.~Geschke.
\newblock Building {EORA}: a global multi-region input--output database at high
  country and sector resolution.
\newblock {\em Economic Systems Research}, 25(1):20--49, 2013.

\bibitem{Leontief1}
W.~Leontief.
\newblock "the economy as a circular flow” (in german).
\newblock {\em Archiv für Sozialwissenschaft und Sozialpolitik}, 60:577--623,
  1928.

\bibitem{Leontief2}
W.~Leontief.
\newblock Quantitative input-output relations in the economic system of the
  united states.
\newblock {\em Review of Economics and Statistics}, 18:105--125, 1936.

\bibitem{melgar2022}
S.~G. Melgar and J.~M. And{\'u}jar~M{\'a}rquez.
\newblock New research trends and topics for achieving energy efficiency in
  buildings: Both new and rehabilitated.
\newblock {\em Energies}, 15(3):851, 2022.

\bibitem{Miller}
R.E. Miller and P.~D. Blair.
\newblock {\em Input-output analysis: foundations and extensions}.
\newblock Cambridge University Press, 2009.

\bibitem{Christou2021}
C.~Odysseas.
\newblock Energy security in turbulent times towards the {E}uropean {G}reen
  {D}eal.
\newblock {\em Politics and Governance}, 9:360–369, 2021.

\bibitem{OLATOMIWA2021}
L.~Olatomiwa, S.~Mekhilef, M.S. Ismail, and M.~Moghavvemi.
\newblock Energy management strategies in hybrid renewable energy systems: {A}
  review.
\newblock {\em Renewable and {S}ustainable {E}nergy {R}eviews}, 62:821--835,
  2016.

\bibitem{qier}
A.~Qier, A.~Haizhong, F.~Wei, and W.~Lang.
\newblock Embodied energy flow network of {C}hinese industries: a complex
  network theory based analysis.
\newblock {\em Energy Procedia}, 61:369--372, 2014.

\bibitem{owidenergy}
H.~Ritchie and M.~Roser.
\newblock Energy.
\newblock {\em Our World in Data}, 2020.
\newblock https://ourworldindata.org/energy.

\bibitem{Serrano2007}
M~Angeles Serrano, Mari{\'a}n Bogu{\~n}{\'a}, and Alessandro Vespignani.
\newblock Patterns of dominant flows in the world trade web.
\newblock {\em Journal of Economic Interaction and Coordination},
  2(2):111--124, 2007.

\bibitem{shi2017}
J.~Shi, H.~Li, J.~Guan, X.~Sun, Q.~Guan, and X.~Liu.
\newblock Evolutionary features of global embodied energy flow between sectors:
  a complex network approach.
\newblock {\em Energy}, 140:395--405, 2017.

\bibitem{Sun}
X.~Sun, J.~Li, H.~Qiao, and Zhang B.
\newblock Energy implications of {C}hina's regional development: new insights
  from multi-regional input-output analysis.
\newblock {\em Applied Energy}, 196:118--131, 2017.

\bibitem{tzekina2008}
I.~Tzekina, K.~Danthi, and D.~N. Rockmore.
\newblock Evolution of community structure in the world trade web.
\newblock {\em The European Physical Journal B}, 63(4):541--545, 2008.

\bibitem{WANG2019}
X.~Wang, J.~J. Klemeš, X.~Dong, W.~Fan, Z.~Xu, Y.~Wang, and P.~Sabev~Varbanov.
\newblock Air pollution terrain nexus: A review considering energy generation
  and consumption.
\newblock {\em Renewable and Sustainable Energy Reviews}, 105:71--85, 2019.

\bibitem{Wiedmann}
T.~Wiedmann, M.~Lenzen, K.~Turner, and J.~Barrett.
\newblock Examining the global environmental impact of regional consumption
  activities—part 2: Review of input-output models for the assessment of
  environmental impacts embodied in trade.
\newblock {\em Ecological Economics}, 61:15--26, 2007.

\bibitem{Wiedmann2}
T.~Wiedmann, R.~Wood, M.~Lenzen, J.~Minx, D.~Guan, and J.~Barrett.
\newblock A carbon footprint time series of the {UK} – results from a
  multi-region input-output model.
\newblock {\em Economic Systems Research}, 22:19--42, 2010.

\bibitem{WU2016}
X.D. Wu, X.H. Xia, G.Q. Chen, X.F. Wu, and B.~Chen.
\newblock Embodied energy analysis for coal-based power generation
  system-highlighting the role of indirect energy cost.
\newblock {\em Applied Energy}, 184:936--950, 2016.

\bibitem{su11154147}
A.~Zaharia, M.~C. Diaconeasa, L.~Brad, G.-R. Lădaru, and C.~Ioanăș.
\newblock Factors influencing energy consumption in the context of sustainable
  development.
\newblock {\em Sustainability}, 11(15), 2019.

\bibitem{Zhao2020}
N.~Zhao and F.ù You.
\newblock Can renewable generation, energy storage and energy efficient
  technologies enable carbon neutral energy transition?
\newblock {\em Applied Energy}, 279:115889, 2020.

\bibitem{Zhong}
W.~Zhong, H.~An, L.~Shen, W.~Fang, X.~Gao, and D.~Dong.
\newblock The roles of countries in the international fossil fuel trade: an
  emergy and network analysis.
\newblock {\em Energy Policy}, 100:365--376,, 2017.

\bibitem{en14175268}
Y.~L. Zhukovskiy, D.~E. Batueva, A.~D. Buldysko, B.~Gil, and V.~V. Starshaia.
\newblock Fossil energy in the framework of sustainable development: Analysis
  of prospects and development of forecast scenarios.
\newblock {\em Energies}, 14(17), 2021.

\end{thebibliography}

\end{document}